\documentclass[aps,prl,twocolumn,showpacs,nofootinbib]{revtex4-1}

\usepackage{amsmath,amssymb,bm}
\usepackage[dvips]{graphicx}
\usepackage{hyperref}
\usepackage{yfonts}
\usepackage{color}
\newcommand{\beq}{\begin{eqnarray}}
\newcommand{\eeq}{\end{eqnarray}}
\newcommand{\SU}{\text{SU}}
\newcommand{\U}{\text{U}}
\renewcommand\d{\partial}
\begin{document}

\title{Chiral Alfv\'en Wave in Anomalous Hydrodynamics}

\author{Naoki Yamamoto}
\affiliation{Department of Physics, Keio University,
Yokohama 223-8522, Japan}

\begin{abstract}
We study the hydrodynamic regime of chiral plasmas at high temperature. 
We find a new type of gapless collective excitation induced by chiral effects 
in an external magnetic field.  This is a transverse wave, and it is present even 
in incompressible fluids, unlike the chiral magnetic and chiral vortical waves. 
The velocity is proportional to the coefficient of the gravitational anomaly.
We briefly discuss the possible relevance of this ``chiral Alfv\'en wave'' in 
physical systems.
\end{abstract}
\pacs{11.15.-q, 47.75.+f, 52.35.Bj}
\maketitle

\emph{Introduction.}---%
Hot and/or dense matter with chirality imbalance is considered to be realized in a 
broad range of systems, including quark-gluon plasmas produced in heavy-ion collisions 
\cite{Kharzeev:2007jp,Fukushima:2008xe}, the early Universe \cite{Joyce:1997uy,Boyarsky:2011uy}, 
compact stars and supernovae \cite{Charbonneau:2009ax,Ohnishi:2014uea}, and 
Weyl semimetals \cite{Vishwanath,BurkovBalents,Xu-chern}. The main reason 
why such chiral matter has attracted much attention recently is that it exhibits 
anomalous transport phenomena, related to quantum anomalies in field theory \cite{Adler,BellJackiw}. 
Two prominent examples are the so-called chiral magnetic effect (CME) 
\cite{Vilenkin:1980fu,Nielsen:1983rb,Alekseev:1998ds,Fukushima:2008xe} 
and chiral vortical effect (CVE) \cite{Vilenkin:1979ui,Erdmenger:2008rm,Banerjee:2008th,
Son:2009tf,Landsteiner:2011cp}, which are the currents along the direction of a 
magnetic field and vorticity in chiral matter.

On the theoretical side, chiral (or anomalous) hydrodynamics \cite{Son:2009tf} and chiral kinetic theory 
\cite{Son:2012wh,Stephanov:2012ki,Chen:2012ca,Manuel:2014dza}, which describe 
anomalous transport phenomena in chiral plasmas, have been formulated.
The chiral plasmas have also been found to exhibit a new type of density wave in an external 
magnetic field or with a vorticity, called the chiral magnetic wave (CMW) \cite{Newman:2005hd,Kharzeev:2010gd} 
and the chiral vortical wave (CVW) \cite{Jiang:2015cva}, as well as the unstable collective mode
in the presence of dynamical (color) electromagnetic fields, called the chiral plasma instability 
(CPI) \cite{Akamatsu:2013pjd,Khaidukov:2013sja} (see also Refs.~\cite{Joyce:1997uy,Boyarsky:2011uy}). 
For recent numerical and analytical applications of chiral (magneto)hydrodynamics, see 
Ref.~\cite{Hongo:2013cqa} and Refs.~\cite{Giovannini:2013oga,Boyarsky:2015faa}, respectively.

In this paper, we show that there exists a new type of gapless collective excitation 
specific for charged chiral plasmas in an external magnetic field. Unlike the CMW and 
CVW, this is a transverse wave, and it is present even in incompressible fluids. This is 
somewhat similar to the Alfv\'en wave in normal charged plasmas, which propagates 
without compressing the medium, driven by magnetic tension forces \cite{Biskamp}. 
As we will show, the velocity of this new mode is proportional to the coefficient of the mixed 
gauge-gravitational anomaly for chiral fermions. We call it the ``chiral Alfv\'en wave" (CAW). 
Since the CAW is the first example of the transverse wave with a vector-type perturbation 
induced by chiral effects, it should provide a unique signature in physical observables. 
We briefly discuss its possible phenomenological implications.
Throughout the paper, we set $\hbar=c=e=1$.

\emph{Physical argument.}---%
Before going to a mathematical analysis based on chiral hydrodynamic equations, 
we first provide a physical argument as to why a new type of wave
(which is different from the CMW and CVW) can exist for chiral fluids in 
an external magnetic field. We consider a fluid of single right-handed chiral 
fermions at high temperature $T$ and zero chemical potential $\mu=0$.

\begin{figure}[b]
\begin{center}
\includegraphics[width=5cm]{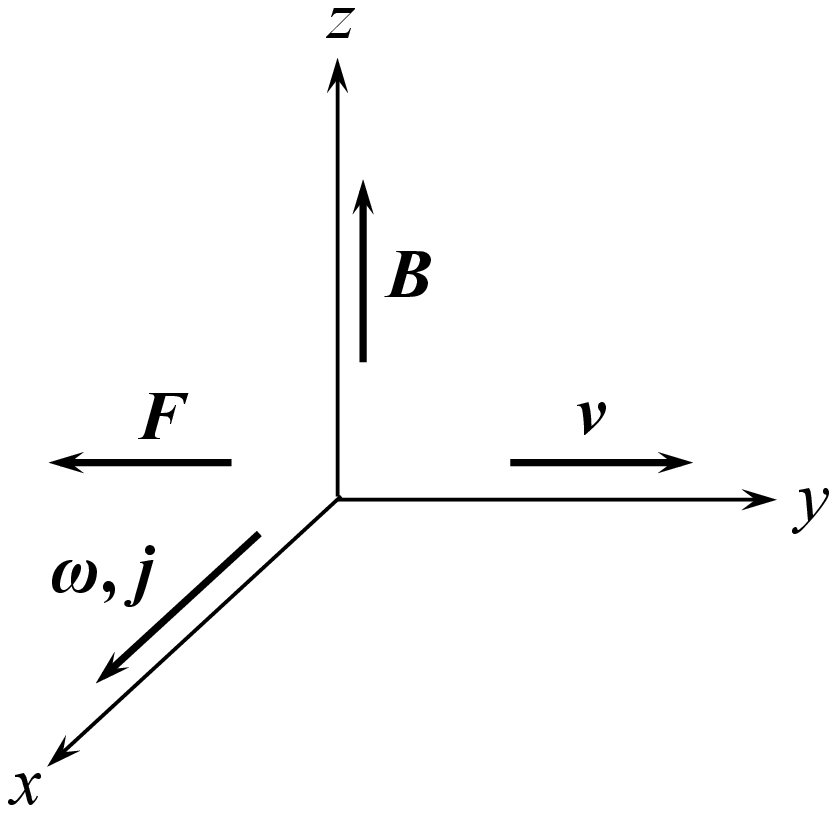}
\end{center}
\vspace{-0.7cm}
\caption{Configuration of the vectors, ${\bm B}$, ${\bm v}$, ${\bm \omega}$, 
${\bm j}$, and ${\bm F}$.}
\label{fig:CAW}
\end{figure}

As shown in Fig.~\ref{fig:CAW}, we take the external magnetic field ${\bm B}$ in 
the positive $z$ direction and consider the perturbation of the local fluid velocity 
${\bm v}$ in the positive $y$ direction, ${\bm v}=v(z) \hat {\bm y}$ with $\d_z v(z)<0$. 
For this local fluid velocity ${\bm v}$, the vorticity ${\bm \omega} = {\bm \nabla} \times {\bm v}$ 
points in the positive $x$ direction. In the chiral fluid, the vorticity induces the local chiral 
vortical current, ${\bm j} \propto T^2{\bm \omega}$ \cite{Vilenkin:1979ui,Landsteiner:2011cp}. 
This current then receives the Lorentz force, ${\bm F} = {\bm j} \times {\bm B}$, in 
the negative $y$ direction. Notice that this is the opposite direction as the original
fluid velocity ${\bm v}$, so the Lorentz force acts as a restoring force; it makes the 
perturbation of the fluid velocity ${\bm v}$ oscillate, and this is the origin of a wave.

This argument holds even when the fluids are incompressible like the Alfv\'en 
wave in the normal charged fluids. This should be contrasted with sound waves 
in normal fluids and the CMW and CVW in chiral fluids, which can propagate 
only in compressible fluids. Apparently, the CVE specific for chiral fluids is 
essential for the presence of this wave---thus, the name chiral Alfv\'en wave. 

In the following, we put this argument on a formal mathematical basis and derive 
its wave equation together with the explicit expression for the velocity of the CAW.

\emph{Chiral hydrodynamics.}---%
Let us start with the generic hydrodynamics for plasmas of single right-handed 
chiral fermions in {\it external} electromagnetic fields. 
The hydrodynamic equations read \cite{Son:2009tf}%
\footnote{In this paper, we use the ``mostly minus" metric signature 
$g^{\mu \nu} = {\rm diag}(1,-1,-1,-1)$.}
\begin{gather}
\label{dT}
\d_{\mu}T^{\mu\nu} =F^{\nu\lambda}j_{\lambda}, \\
\label{dj}
\d_{\mu}j^{\mu} = - C E^{\mu} B_{\mu}.
\end{gather}
Here $T^{\mu \nu}$ is the energy-momentum tensor, $j^{\mu}$ is the electric current, 
$F^{\mu \nu}$ is the field strength, $C$ is the anomaly coefficient [see Eq.~(\ref{C}) below], 
$E^{\mu}=F^{\mu \nu} u_{\nu}$, and $B^{\mu}=\frac{1}{2}\epsilon^{\mu \nu \alpha \beta}u_{\nu}F_{\alpha \beta}$,
with $u^{\mu}=\gamma(1, {\bm v})$ being the local fluid velocity. 

In the Landau-Lifshitz frame, $T^{\mu \nu}$ and $j^{\mu}$ are given by \cite{Son:2009tf}
\begin{gather}
T^{\mu \nu}=(\epsilon+P)u^{\mu}u^{\nu}-Pg^{\mu \nu}+\tau^{\mu \nu}, \\
j^{\mu}=nu^{\mu} + \xi \omega^{\mu} + \xi_B B^{\mu} + \nu^{\mu},
\end{gather} 
with $\epsilon$ the energy density, $P$ the pressure, $n$ the charge density, 
and $\omega^{\mu}=\epsilon^{\mu \nu \alpha \beta} u_{\nu} \d_{\alpha} u_{\beta}$
the vorticity.%
\footnote{Note that the definition of the vorticity $\omega^{\mu}$ here is different 
from the one in Ref.~\cite{Son:2009tf} by a factor of 2.} 
The dissipative effects, such as the conductivity $\sigma$ and viscosities, are 
incorporated in $\nu^{\mu}$ and $\tau^{\mu \nu}$, which we ignore to simplify 
our argument for a moment. (We discuss these corrections later.)
The transport coefficients $\xi_B$ and $\xi$, corresponding to the CME 
\cite{Vilenkin:1980fu,Nielsen:1983rb,Alekseev:1998ds,Fukushima:2008xe} 
and CVE \cite{Vilenkin:1979ui,Erdmenger:2008rm,Banerjee:2008th,Son:2009tf,Landsteiner:2011cp}, 
take the form \cite{Son:2009tf,Neiman:2010zi}
\begin{gather}
\xi_B = C \mu \left(1-\frac{1}{2} \frac{n\mu}{\epsilon+P} \right)
- \frac{D}{2}\frac{nT^2}{\epsilon+P} \,, \\
\xi = \frac{C}{2} \mu^2 \left(1-\frac{2}{3} \frac{n\mu}{\epsilon+P} \right) 
+ \frac{D}{2} T^2 \left(1- \frac{2 n \mu }{\epsilon+P} \right) \,,
\end{gather}
where $T$ is the temperature and $\mu$ is the chemical potential.
The transport coefficients $C$ and $D$ are related to the chiral anomaly 
and mixed gauge-gravitational anomaly, respectively, 
as \cite{Son:2009tf,Landsteiner:2011cp,Gao:2012ix,Golkar:2012kb,Jensen:2012kj}
\begin{align}
\label{C}
C=\frac{1}{4\pi^2}, \qquad
D=\frac{1}{12}.
\end{align}

We now explicitly write down the hydrodynamic equations in an external magnetic field. 
We assume the bulk collective flow to be nonrelativistic, $|{\bm v}|\ll 1$, 
despite individual constituents of the fluid being relativistic. 
To be specific, we consider the following counting scheme: 
$\d_t \sim O(\epsilon_t)$, ${\bm \nabla} \sim O(\epsilon_s)$, and ${\bm v} \sim O(\delta)$
with three independent expansion parameters, $\epsilon_{s,t} \ll 1$ and $\delta \ll 1$. 
We take the gauge field $A^{\mu}$ to be the same order as $T$ and $\mu$, 
so that ${\bm B} \sim O(\epsilon_s)$.

As usual, it is convenient to consider the longitudinal and transverse projections of 
Eq.~(\ref{dT}) with respect to the fluid velocity \cite{Landau}, 
\begin{gather}
\label{hydro_L}
u_{\nu}\d_{\mu}T^{\mu \nu} = u_{\nu}F^{\nu \lambda} j_{\lambda}, \\
\label{hydro_T}
(g^{\rho}_{\nu}-u^{\rho}u_{\nu})\d_{\mu}T^{\mu \nu} = (g^{\rho}_{\nu}-u^{\rho}u_{\nu})F^{\nu \lambda} j_{\lambda}.
\end{gather}
Although hydrodynamic equations to the linear order in ${\bm v}$ will be sufficient for 
the purpose of deriving the wave equation of the CAW, we first write down a more 
generic ``Euler equation" taking into account chiral effects. To this end, we keep the 
terms to the order of $O(\epsilon_t \delta, \epsilon_s \delta^2, \epsilon_s^2 \delta)$. 
Then, Eqs.~(\ref{hydro_L}), (\ref{hydro_T}) for the temporal and spatial components ($\rho=0,i$), 
and (\ref{dj}) are given by
\begin{gather}
\label{E}
(\d_t + {\bm v} \cdot {\bm \nabla}){\epsilon}
+(\epsilon+P){\bm \nabla} \cdot {\bm v} = 0,
\\
\label{P}
({\bm v} \cdot {\bm \nabla})P = 0, 
\\
\label{v}
(\epsilon+P)(\d_t + {\bm v} \cdot {\bm \nabla}){\bm v}
= -{\bm \nabla}P-{\bm v}(\d_t + {\bm v} \cdot {\bm \nabla}) P + {\bm j} \times {\bm B}, 
\\
\label{n}
\d_t n + {\bm \nabla} \cdot {\bm j} = 0,
\end{gather}
respectively. Here
\beq
\label{j}
{\bm j}=n {\bm v} + \xi {\bm \omega} + \xi_B {\bm B},
\eeq
with ${\bm \omega}={\bm \nabla} \times {\bm v}$. 
The terms ${\bm v}\cdot({\bm j} \times {\bm B})$ on the right-hand sides of 
Eqs.~(\ref{E}) and (\ref{P}) are of order $O(\epsilon_s^2 \delta^2)$ and are ignored. 

For plasmas with homogeneous and static $\epsilon$, $P$, and $n$
(which is the case where the variations of $T$ and $\mu$ are much smaller and slower 
than ${\bm v}$), the hydrodynamic equations are further simplified to
\begin{gather}
\label{hydro}
(\epsilon+P)(\d_t {\bm v} + {\bm v} \cdot {\bm \nabla}{\bm v}) = (n {\bm v} + \xi {\bm \omega}) \times {\bm B}, \\
\label{incomp}
{\bm \nabla} \cdot {\bm v} = 0.
\end{gather}
Equation (\ref{incomp}) is the incompressibility condition of fluids.

\emph{Chiral Alfv\'en wave.}---%
We now assume homogeneous plasmas at high temperature $T \gg \mu$, where 
we can ignore the contribution of $\mu$ (and so $n$) \cite{note}. Let us consider a small 
perturbation of ${\bm v}$, similarly to the analysis in sound waves in hydrodynamics 
and normal Alfv\'en waves in magnetohydrodynamics \cite{Biskamp}. 
The linearized chiral hydrodynamic equation (\ref{hydro}) in a magnetic field 
[to the order of $O(\epsilon_t \delta, \epsilon_s^2 \delta)$ by assuming 
$\delta \ll \epsilon_{s,t}$] is given by
\beq
\label{v1}
(\epsilon + P)\d_t {\bm v} = \xi {\bm \omega} \times {\bm B},
\eeq
where $\xi = DT^2/2$. Using
${\bm \omega} \times {\bm B}=({\bm B}\cdot {\bm \nabla}){\bm v}-{\bm \nabla}({\bm B}\cdot {\bm v})$,
the above equation can be rewritten as
\beq
\label{v2}
(\epsilon + P)\d_t {\bm v} = \xi [({\bm B}\cdot {\bm \nabla}){\bm v}-{\bm \nabla}({\bm B}\cdot {\bm v})].
\eeq
Taking the divergence of Eq.~(\ref{v2}) and using Eq.~(\ref{incomp}), we have
${\bm \nabla}^2 ({\bm B}\cdot {\bm v})=0$.
To satisfy this equation, we set
\beq
{\bm B}\cdot {\bm v}=0,
\eeq
i.e., the direction of ${\bm v}$ is taken to be perpendicular to ${\bm B}$.

Without loss of generality, we take the magnetic field in the $z$ direction,  
${\bm B}=B {\hat {\bm z}}$. From Eq.~(\ref{v2}), we then obtain the wave equation
\beq
\label{v_linear}
\d_t {\bm v} = V_{\rm T} \d_z {\bm v},
\eeq
where
\beq
\label{V_T}
V_{\rm T} 
= \frac{D}{2} \frac{T^2}{\epsilon + P}B.
\eeq

Substituting the plane-wave solution of the form
${\bm v}={\bm v}_0 e^{-i\omega t + i{\bm k}\cdot{\bm x}}$, we get 
the dispersion relation
\beq
\label{dispersion}
\omega = -V_{\rm T} k_z,
\eeq
so the wave propagates in the opposite direction as the magnetic field.
(For the fluid with left-handed chiral fermions, the wave propagates with the 
velocity $V_T$ in the same direction as the magnetic field.)
As the velocity of this collective mode is proportional to the coefficient $D$
related to the mixed gauge-gravitational anomaly for chiral fermions, this 
phenomenon is specific for chiral fluids, and it is not present in normal charged 
fluids. This is the CAW. The incompressibility condition (\ref{incomp}) means 
that the CAW is a transverse wave, ${\bm k} \cdot {\bm v} = 0$.

This clarifies the difference between the CAW and CMW, the latter of which 
is also a gapless collective excitation in chiral fluids with a magnetic field; 
the CMW is a longitudinal density wave and appears only in compressible 
fluids ($\d_t n \neq 0$) \cite{Newman:2005hd, Kharzeev:2010gd}, while 
the CAW is present even in incompressible fluids ($\d_t n =0$). 
Note also that the conventional Alfv\'en wave appears when the magnetic fields 
are dynamical \cite{Biskamp}. In contrast, the CAW exists even when the 
magnetic fields are external.

For high-temperature chiral plasma, $\epsilon, P \propto T^4$, and
the velocity of the CAW, given by Eq.~(\ref{V_T}), is $V_{\rm T} \propto B/T^2$.
On the other hand, the velocity of the CMW, given by 
$V_{\rm L} \equiv B/(4\pi^2 \chi)$ with 
$\chi \equiv \d n/\d \mu$ the susceptibility \cite{Kharzeev:2010gd}, is also
$V_{\rm L} \propto B/T^2$. Hence, $V_{\rm T}$ and $V_{\rm L}$ coincide
up to (generally different) prefactors. Note that $V_{\rm T,L} \ll 1$ in our 
counting scheme, where ${\bm B} \sim O(\epsilon_s)$. 

When the dissipative effects are included, the terms $\nu {\bm \nabla}^2 {\bm v}$
and $-{\tilde \sigma} B^2 {\bm v}$, with $\nu$ the kinematic viscosity and 
${\tilde \sigma}=\sigma/(\epsilon + P)$, are also added to the right-hand side of 
Eq.~(\ref{v_linear}). Then, the dispersion relation of Eq.~(\ref{dispersion}) becomes
\beq
\omega = -V_{\rm T} k_z - i (\nu k^2 + \tilde \sigma B^2)
\eeq
with $k \equiv |{\bm k}|$.
In this case, the CAW propagates with the velocity $V_{\rm T}$, but is damped 
by dissipation.

\emph{Discussions.}---%
Let us discuss phenomenological implications of the CAW. One possible physical situation 
is the chiral fluid in the early Universe. Above the electroweak phase transition temperature 
where the $\SU(2)_{\rm L} \times \U(1)_{\rm Y}$ symmetry is restored, the massless Abelian 
gauge field that can propagate at long distances is the hypermagnetic field associated 
with the $\U(1)_{\rm Y}$ hypercharge, rather than the ordinary magnetic field associated with 
the $\U(1)_{\rm EM}$ charge.  As the hypermagnetic field couples differently to right- and 
left-handed electrons, the high-temperature plasma there is chiral \cite{Giovannini:1997eg}. 

One expects the emergence of the CAW in such chiral fluids with strong hypermagnetic 
fields, which should affect the polarization anisotropies of the cosmic microwave background radiation (CMBR).
Since the CAW is the vector-type perturbation (i.e., the velocity $\delta {\bm v}$) while the 
CMW and CVW are scalar-type ones (i.e., the density $\delta n$), the former should leave a 
peculiar signature that can be distinguished from the latter, e.g., parity-odd correlations of 
multipole amplitudes with different angular momenta in the CMBR. 
We defer this question to future work.

It would be interesting to study the possible roles of the CAW in the turbulence of 
chiral magnetohydrodynamics relevant to the evolution of the primordial magnetic field of 
the Universe. One can also consider the analogue of the CAW in rotating chiral fluids 
(without external electromagnetic fields), which is relevant to the physics of hot neutrino gas.

Finally, the CAW revealed in this paper may be understood in the language of chiral
kinetic theory \cite{Son:2012wh, Stephanov:2012ki, Chen:2012ca, Manuel:2014dza}, 
in a way similar to Refs.~\cite{Stephanov:2014dma, Jiang:2015cva}. 


\acknowledgments
We thank Y.~Akamatsu, M.~Hongo, X.~G.~Huang, and Y.~Yin for useful comments.
This work was supported by JSPS KAKENHI Grant No. 26887032.


\begin{thebibliography}{99}

  \bibitem{Kharzeev:2007jp}
  D.~E.~Kharzeev, L.~D.~McLerran, and H.~J.~Warringa,
  Nucl.\ Phys.\ A {\bf 803}, 227 (2008).
  
    \bibitem{Fukushima:2008xe}
  K.~Fukushima, D.~E.~Kharzeev, and H.~J.~Warringa,
  Phys.\ Rev.\ D {\bf 78}, 074033 (2008).

  \bibitem{Joyce:1997uy}
  M.~Joyce and M.~E.~Shaposhnikov,
  Phys.\ Rev.\ Lett.\  {\bf 79}, 1193 (1997).

  \bibitem{Boyarsky:2011uy} 
  A.~Boyarsky, J.~Frohlich, and O.~Ruchayskiy,
  Phys.\ Rev.\ Lett.\  {\bf 108}, 031301 (2012).  
  
    \bibitem{Charbonneau:2009ax}
  J.~Charbonneau and A.~Zhitnitsky,
  JCAP {\bf 1008}, 010 (2010).

  \bibitem{Ohnishi:2014uea} 
  A.~Ohnishi and N.~Yamamoto,
  arXiv:1402.4760 [astro-ph.HE].

  \bibitem{Vishwanath}
  X.~Wan, A.~M.~Turner, A.~Vishwanath, and S.~Y.~Sav\-ra\-sov,
  Phys.\ Rev.\ B {\bf 83}, 205101 (2011).

\bibitem{BurkovBalents}
  A.~A.~Burkov and L.~Balents,
  Phys.\ Rev.\ Lett.\ {\bf 107}, 127205 (2011).

\bibitem{Xu-chern}
  G.~Xu, H.~Weng, Z.~Wang, X.~Dai, and Z.~Fang,
  Phys.\ Rev.\ Lett. {\bf 107}, 186806 (2011).
  
  \bibitem{Adler}
  S.~Adler,
  Phys.\ Rev.\ {\bf 177}, 2426 (1969).

  \bibitem{BellJackiw}
  J.~S.~Bell and R.~Jackiw,
  Nuovo Cimento {\bf 60A}, 47 (1969).

  \bibitem{Vilenkin:1980fu}
  A.~Vilenkin,
  Phys.\ Rev.\ D {\bf 22}, 3080 (1980).
  
  \bibitem{Nielsen:1983rb} 
  H.~B.~Nielsen and M.~Ninomiya,
  Phys.\ Lett.\ B {\bf 130}, 389 (1983).  

  \bibitem{Alekseev:1998ds} 
  A.~Y.~Alekseev, V.~V.~Cheianov, and J.~Frohlich,
  Phys.\ Rev.\ Lett.\  {\bf 81}, 3503 (1998).  
  
  \bibitem{Vilenkin:1979ui} 
  A.~Vilenkin,
  Phys.\ Rev.\ D {\bf 20}, 1807 (1979).
  
  \bibitem{Erdmenger:2008rm} 
  J.~Erdmenger, M.~Haack, M.~Kaminski, and A.~Yarom,
  JHEP {\bf 0901}, 055 (2009).
  
  \bibitem{Banerjee:2008th} 
  N.~Banerjee, J.~Bhattacharya, S.~Bhattacharyya, S.~Dutta, R.~Loganayagam, and P.~Surowka,
  JHEP {\bf 1101}, 094 (2011).
  
  \bibitem{Son:2009tf}
  D.~T.~Son and P.~Surowka,
  Phys.\ Rev.\ Lett.\  {\bf 103}, 191601 (2009).  
  
  \bibitem{Landsteiner:2011cp} 
  K.~Landsteiner, E.~Megias, and F.~Pena-Benitez,
  Phys.\ Rev.\ Lett.\  {\bf 107}, 021601 (2011);
  Lect.\ Notes Phys.\  {\bf 871}, 433 (2013).

  \bibitem{Son:2012wh}
  D.~T.~Son and N.~Yamamoto,
  Phys.\ Rev.\ Lett.\  {\bf 109}, 181602 (2012);
  Phys.\  Rev.\ D {\bf 87}, 085016 (2013).

  \bibitem{Stephanov:2012ki} 
  M.~A.~Stephanov and Y.~Yin,
  Phys.\ Rev.\ Lett.\  {\bf 109}, 162001 (2012).

  \bibitem{Chen:2012ca} 
  J.~-W.~Chen, S.~Pu, Q.~Wang, and X.~-N.~Wang,
  Phys.\ Rev.\ Lett.\  {\bf 110}, 262301 (2013).
  

  \bibitem{Manuel:2014dza} 
  C.~Manuel and J.~M.~Torres-Rincon,
  Phys.\ Rev.\ D {\bf 90}, 076007 (2014).
    
  \bibitem{Newman:2005hd}
  G.~M.~Newman,
  JHEP {\bf 0601}, 158 (2006).
  
  \bibitem{Kharzeev:2010gd} 
  D.~E.~Kharzeev and H.~U.~Yee,
  Phys.\ Rev.\ D {\bf 83}, 085007 (2011).
  
  \bibitem{Jiang:2015cva} 
  Y.~Jiang, X.~G.~Huang, and J.~Liao,
  arXiv:1504.03201 [hep-ph].
  
  \bibitem{Akamatsu:2013pjd}
  Y.~Akamatsu and N.~Yamamoto,
  Phys.\ Rev.\ Lett.\  {\bf 111}, 052002 (2013); 
  Phys.\ Rev.\ D {\bf 90}, 125031 (2014).
  
  \bibitem{Khaidukov:2013sja} 
  Z.~V.~Khaidukov, V.~P.~Kirilin, A.~V.~Sadofyev, and V.~I.~Zakharov,
  arXiv:1307.0138 [hep-th].
  
  \bibitem{Hongo:2013cqa} 
  M.~Hongo, Y.~Hirono, and T.~Hirano,
  arXiv:1309.2823 [nucl-th];
  Y.~Hirono, T.~Hirano, and D.~E.~Kharzeev,
  arXiv:1412.0311 [hep-ph].
  
  \bibitem{Giovannini:2013oga}
  M.~Giovannini,
  Phys.\ Rev.\ D {\bf 88}, 063536 (2013).
  
  \bibitem{Boyarsky:2015faa} 
  A.~Boyarsky, J.~Frohlich, and O.~Ruchayskiy,
  Phys.\ Rev.\ D {\bf 92}, 043004 (2015).
  
  \bibitem{Biskamp}
  See, e.g., D.~Biskamp, {\it Nonlinear Magnetohydrodynamics}
  (Cambridge University Press, Cambridge, England, 1993).
  
  
  \bibitem{Neiman:2010zi} 
  Y.~Neiman and Y.~Oz,
  JHEP {\bf 1103}, 023 (2011).
  
  \bibitem{Gao:2012ix} 
  J.~H.~Gao, Z.~T.~Liang, S.~Pu, Q.~Wang, and X.~N.~Wang,
  Phys.\ Rev.\ Lett.\  {\bf 109}, 232301 (2012).
  
  \bibitem{Golkar:2012kb}
  S.~Golkar and D.~T.~Son,
  JHEP {\bf 1502}, 169 (2015).
  
  \bibitem{Jensen:2012kj} 
  K.~Jensen, R.~Loganayagam, and A.~Yarom,
  JHEP {\bf 1302}, 088 (2013).
  
  \bibitem{Landau}
  L.~D.~Landau and E.~M.~Lifshitz,
  {\it Fluid Mechanics} (Pergamon, New York, 1959).
  
  \bibitem{note}
  When the contribution of the first term on the right-hand side of Eq.~(\ref{hydro}) is dominant 
  compared with the second one, the dispersion relation in Eq.~(\ref{dispersion})
  becomes $\omega = \pm n B/(\epsilon + P)$, corresponding to the Larmor frequency.
  
  \bibitem{Giovannini:1997eg} 
  M.~Giovannini and M.~E.~Shaposhnikov,
  Phys.\ Rev.\ D {\bf 57}, 2186 (1998).
  
  \bibitem{Stephanov:2014dma} 
  M.~Stephanov, H.~U.~Yee, and Y.~Yin,
  Phys.\ Rev.\ D {\bf 91}, 125014 (2015).

\end{thebibliography}
\end{document}